Article

# Ambivalent and Consistent Relationships: The Role of Personal Networks in Cases of Domestic Violence

Elisa Bellotti [1],*, Susanne Boethius [2], Malin Åkerström [2] and Margareta Hydén [3]

[1] Mitchell Centre for Social Network Analysis, School of Social Sciences, University of Manchester, UK; E-Mail: elisa.bellotti@manchester.ac.uk

[2] Department of Sociology, Lund University, Sweden; E-Mails: susanne.boethius@soc.lu.se (S.B.), malin.akerstrom@soc.lu.se (M.A.)

[3] Department of Culture and Society, Linköping University, Sweden; E-Mail: margareta.hyden@liu.se

* Corresponding author

**Abstract**

Social networks are usually considered as positive sources of social support, a role which has been extensively studied in the context of domestic violence. To victims of abuse, social networks often provide initial emotional and practical help as well useful information ahead of formal institutions. Recently, however, attention has been paid to the negative responses of social networks. In this article, we advance the theoretical debate on social networks as a source of social support by moving beyond the distinction between positive and negative ties. We do so by proposing the concepts of relational ambivalence and consistency, which describe the interactive processes by which people, intentionally or inadvertently, disregard—or align with—each other's role-relational expectations, therefore undermining—or reinforcing—individual's choices of action. We analyse the qualitative accounts of 19 female victims of domestic violence in Sweden, who described the responses of their personal networks during and after the abuse. We observe how the relationships embedded in these networks were described in ambivalent and consistent terms, and how they played a role in supporting or undermining women in reframing their loving relationships as abusive; in accounting or dismissing perpetrators' responsibilities for the abuse; in relieving women from role-expectations and obligations or in burdening them with further responsibilities; and in supporting or challenging their pathways out of domestic abuse. Our analysis suggests that social isolation cannot be considered a simple result of a lack of support but of the complex dynamics in which support is offered and accepted or withdrawn and refused.

**Issue**

This article is part of the issue "In Good Company? Personal Relationships, Network Embeddedness and Social Inclusion" edited by Miranda J. Lubbers (Autonomous University of Barcelona, Spain).

**1. Introduction**

Domestic violence is a crime that affects a considerable number of women, and increasingly men (Black et al., 2008). It touches people of all classes, ages, and ethnicities (Montalvo-Liendo, 2009; Reid, 1993; Sylaska & Edwards, 2014), and it manifests in various forms: stalking, psychological abuse, sexual abuse, physical abuse, and in some cases, homicide. As domestic violence is highly under-reported to institutions, it has often been described as a hidden crime (Novisky & Peralta, 2015), but this does not mean that it happens in a social vacuum. The network of people surrounding the victim is frequently aware or at least suspects that something is wrong (Hydén, 2015). Disclosure to family, friends, colleagues, neighbours, or even acquaintances often constitute the first important step that victims take toward escaping the abuse (Boethius & Åkerström, 2020). Social networks around victims are also often a vital source of social support: people may try to stop the abuse, may help the victim by offering means of escape (especially when the perpetrator controls the victim's access to finance and communication), or they may help report the violence to the authorities (Goodman & Smyth, 2011).

This is why social support has been extensively studied in the context of domestic violence, especially focusing on the availability of support (Carlson et al., 2002). In particular, Coohey (2007) notes that if we measure the size of the network of received support, abused women are not more socially isolated than non-abused women—but if we measure the size of perceived support, abused women appear more socially isolated. The focus on perception suggests an important element: support is related to expectations, as victims may feel that people around them are not doing what they need, or not enough, or not at the right time. People can withdraw help, criticise the victim, become scared, or side with the perpetrator; they can also become a further burden for the victim, who may be worried that by disclosing the violence and asking for help, the perpetrator could retaliate against her or her network. This double role of social networks in abusive relationships has been recently recognised by a growing literature that focuses on the negative aspects of social support (Goodkind et al., 2003; Kocot & Goodman, 2003; Latta & Goodman, 2011, Levendosky et al., 2004; Trotter & Allen, 2009). The very term "social support" has been put under question because "support connotes images of encouragement and care, suggesting that responses from network members are uniformly positive" (Trotter & Allen, 2009, p. 222); however, networks can have a wide range of social responses, including those perceived as negative (Rook, 1984; Sandler & Barrera, 1984). As social support does not capture the complexities of networks' reactions, scholars have suggested replacing the term with "social reactions" (Trotter & Allen, 2009, p. 228) or "response network" (Hydén, 2016).

In this article, we build on the above theories of social responses by looking at how supportive social networks have been according to the descriptions of 19 female victims of abuse in Sweden. We asked these women to map the personal networks of people who were relevant for them during the abusive relationships and after they decided to report the abuse to the police. We then conducted qualitative interviews in which the relationships were extensively discussed, highlighting their perceived role in helping—or hindering—the victim. The picture emerging from the study shows complex relational dynamics that cannot be reduced to positive or negative responses. People may be supportive, but victims may not recognise the support until they escape from the abuse; they may be supportive immediately after the disclosure of violence, but they may also fade out in time, over burned by the difficulties involved in facing an abusive relationship. They may respond inconsistently to the victim's expectations, by providing support when not required or not expected, or by refusing it when asked. They may side with the perpetrator, blame the victim, dismiss the violence, yet help with childcare, with shelter, and so on. Victims can also hide the abuse, therefore expecting the surrounding network to withdraw, or at least not to interfere, but people may suspect and intervene regardless; alternatively, victims may indirectly disclose the violence, for example, by not hiding bruises, and develop unspoken expectations in the hope that people around them intervene even if not explicitly asked.

To describe and explain such complexities, which may happen within a single relationship but also across the whole network (Trotter & Allen, 2009), and which may unveil over time, we propose the concepts of relational ambivalence and consistency, which describe the relational processes by which people in interactions, intentionally or inadvertently, disregard—or align with—each other's role-relational expectations, therefore undermining—or reinforcing—individual's choices of action. By looking at these alignments and discrepancies, we theoretically redefine the concept of social support by moving beyond the distinction between positive and negative ties, to better understand why ties are perceived as supportive by the victim, and what role they effectively play in helping the victims to escape abusive relationships.

**2. Ambivalence and Consistency: From Roles and Interactions to Relational Definitions**

Sociological ambivalence is a "sensitising concept" (Blumer, 1954; Lüscher, 2002) that has been extensively used to describe conflictual relationships, especially in family and intergenerational relations (Connidis & McMullin, 2002). In its early definitions (Merton, 1976), sociological ambivalence is built and generalised into the structure of social statuses (i.e., gender, class, ethnicity). It refers to incompatible or contradictory normative roles' expectations assigned to a status, for example, the demands required from a doctor who has to be at the same time detached and compassionate, or the conflicts that emerge in a woman who juggles motherhood and work. Merton's sociological definition of ambivalence highlights the relational dimension of norms and counter-norms which define roles expected in social statuses, but it is Bott (1957) who empirically shows how such normative roles are not defined just within a relationship but within the network in which these relationships are embedded. In Bott's analysis, role-relationships are defined as those aspects of a relationship that consist of reciprocal role expectations, and variations in role-relationships are related to the configuration of social networks.

Role theory has subsequently been criticised for the relatively passive image of individuals which downplays their agency in role making (Stryker & Statham, 1985). When people enact their roles (such as that of a mother or of a colleague) they negotiate the norms of interaction (Blumer, 1969), although the type of expectations that individuals experience are constrained by the tendency of humans to typify situations, simplify social contexts, and interpret actions and responses using previous interactions (Blumer, 1969; Mead, 1934; Stryker, 2002). Goffman defines the act of role-performing as scripted in interaction orders which provide a social organisation of experience, or, as he calls it, "social ritualisation" (Goffman, 1983, p. 3). For performances to be consistently supported, they need to at least partially follow expected, culturally accepted interactive orders (Goffman, 1961, 1983). When actions are not supported, we may feel let down or misunderstood by our networks, and conflicts may arise.

Social network theory usually focuses on the positive social support that personal networks offer (Fischer, 1982; House et al., 1988; Wellman & Wortley, 1990) and typically shows how family, friends, and acquaintances assist respondents, connect them to various resources, and contribute to their physical and mental health (Offer & Fischer, 2018). Recently, however, attention has been given to the reasons why people maintain relationships with others that may not support their decisions, may be perceived as difficult, or may not offer any support at all (Offer & Fischer, 2018; Sarazin, 2021). Offer and Fischer (2018), for example, find that close family members are more likely than non-kin to be described as difficult people, as well as giving support that is not reciprocated or overly burdening, but that normative and institutional constraints may force people to retain difficult and demanding people in their networks.

In the context of domestic abuse, difficult relations are common, but so far, they have been described as negative responses of social support (Goodkind et al., 2003; Kocot & Goodman, 2003; Latta & Goodman, 2011, Levendosky et al., 2004; Trotter & Allen, 2009), with little attention being paid to the dynamic interactive nature of social relationships and their embeddedness in more or less constraining social networks. We define relational ambivalence as the relational dynamic in which people fall short of the victim's expectations, either by providing social support when not expected or requested (positive ambivalence) or by denying support when implicitly or explicitly expected and requested (negative ambivalence). Likewise, we define relational consistency as the relational dynamic in which people either provide support when expected (positive consistency) or do not offer support in line with the victim's expectations (negative consistency). The concepts of ambivalence and consistency, more than positive and negative ties, are better suited to illustrate the alignment or discrepancy between role-relational expectations and actual interactions. Note that in our theoretical framework, role expectations may be implicitly assumed and not always intentionally communicated: in the intimacy of a relationship, people may expect their contacts to respond in a certain way by intuitively understanding these expectations (Simmel, 1984/1907). Also, ambivalence and consistency introduce the element of time, where support may withdraw or change depending on decisions that victims take, for example, whether they disclose the abuse, and how these decisions are shared with their contacts and endorsed by them. Finally, ambivalent and consistent relationships can be reinforced or dismissed by the networks in which they are embedded, which may push others to respond to the abuse in a concerted way that conforms to the other members of the network. The concept of relational ambivalence has been previously used in psychology to explore attitudinal and motivational ambivalence toward romantic partners in the framework of attachment theories (Mikulincer et al., 2010) and in the context of migrations to illustrate the emotional work that individuals do when they take into account others' opinions in their decision processes (Palmberger, 2019). To our knowledge, this is the first time relational ambivalence and its counterpart, consistency, have been used to analyse the content and dynamics of social support.

**3. Methods and Materials**

In this study, we collected qualitative data via teller-focused interviews, specifically designed to explore such sensitive and morally questionable topics as violence in close relationships and the unbalanced power dynamics between interviewer and interviewees, who usually have to reposition themselves from a person of low value to one who is valuable to the research (Hydén, 2014).

The interviews explored the relational expectations and practices embedded in the personal networks of 19 female victims of domestic violence in Sweden, who decided to report the perpetrator to the police and terminate the abusive relationship. All interviews were conducted by the second author, who recruited the interviewees from help centres targeting abused women. The interviews were tape-recorded and transcribed verbatim, and all names and places were fully anonymised. Women were primarily Swedish citizens, but five were born in other countries (three of them had moved to Sweden as children, one in her late teens, and one as an adult). As they all spoke fluent Swedish, the interviews were conducted in Swedish, but the accounts related to relationships as well as the overall content of the interviews were summarised in English to make them accessible to the first author. At the time of the interview, the interviewees were aged between 27 and 55 years old, they had been in intimate relationships with the abusive perpetrators for periods between 3 months and 25 years, and they had experienced abuse for different lengths of time before reporting it to the police, with one woman filing a complaint one week after the first violent incident, while another after being abused for 18 years. Twelve women were mothers and had full or partial custody of their children. For five mothers, the abuser was not the children's father. When interviewed, all women had separated from the abusers and lived in new accommodation. Some lived by themselves, some with their children and some had a new cohabiting partner.

We asked these women to name everyone who was around them during the time in which they were involved in abusive relationships and after they decided to call the police. They talked about what had happened in these relationships, why they were perceived as being there, and who the people were. We initially used the ideas of ambivalence and consistency as sensitising concepts, as in constructs that are derived from the participants' accounts and sensitise the researcher to develop theoretical definitions (Blumer, 1954). In our initial analysis, we consistently noticed that participants found it difficult to define their supportive relationships as positive or negative, or even as supportive at all. We used these descriptions to formulate a preliminary empirical construct of ambivalence and its counterpart, consistency, and informed it with existing theories of sociological ambivalence. Armed with our theoretically refined concepts, we then went back to the interviewing materials and categorised all the instances in which women discussed their relationships in (positive or negative) ambivalent or consistent terms, accounting for their nuances and variability. This categorisation was independently assigned by the first and second authors to increase validity, and all the cases in which the categorisation did not coincide were discussed and resolved. Note that not all relationships were discussed in these terms and that the categories of ambivalence and consistence were not used to classify relationships tout court, but the interactive instances that interviewees focused upon to describe why and how in certain cases they felt supported—or not—by the people around them. This means that the same relationship could have been perceived at times as ambivalent and evolve into consistent, or that only some aspects of support were described as ambivalent, while others were perceived consistent, as we will illustrate in the analysis.

From the qualitative materials, we also searched for instances in which the perpetrator actively tries to isolate the victim (for example, by confiscating her phone, by controlling her social media, or by relocating to an unfamiliar neighbourhood), the victim attempts to isolate herself (by refusing to answer calls, by avoiding meeting people, by withdrawing from social activities), and the personal network withdraws itself (by refusing to maintain contact with the victim, by forcing her to choose between them and the perpetrator). During the interview, we asked women to map all the people they named on two targets of concentric circles (Kahn & Antonucci, 1980), representing the time before and after they called the police, where the closer the people were placed to the centre of the target, the closer the relationship was perceived by the interviewees. We coded the type of people named into categories: family, relatives, friends and neighbours, and work and school. For each person, when known, we coded the gender, the age, and if they knew about the abuse before the victim reported the perpetrator to the police or not. Finally, we asked the interviewees to indicate which of these people knew each other, and with this information, we built their personal network structures. Figure 1 exemplifies the two personal networks of one of the interviewees, Louise, during the abusive relationship, and after she called the police.

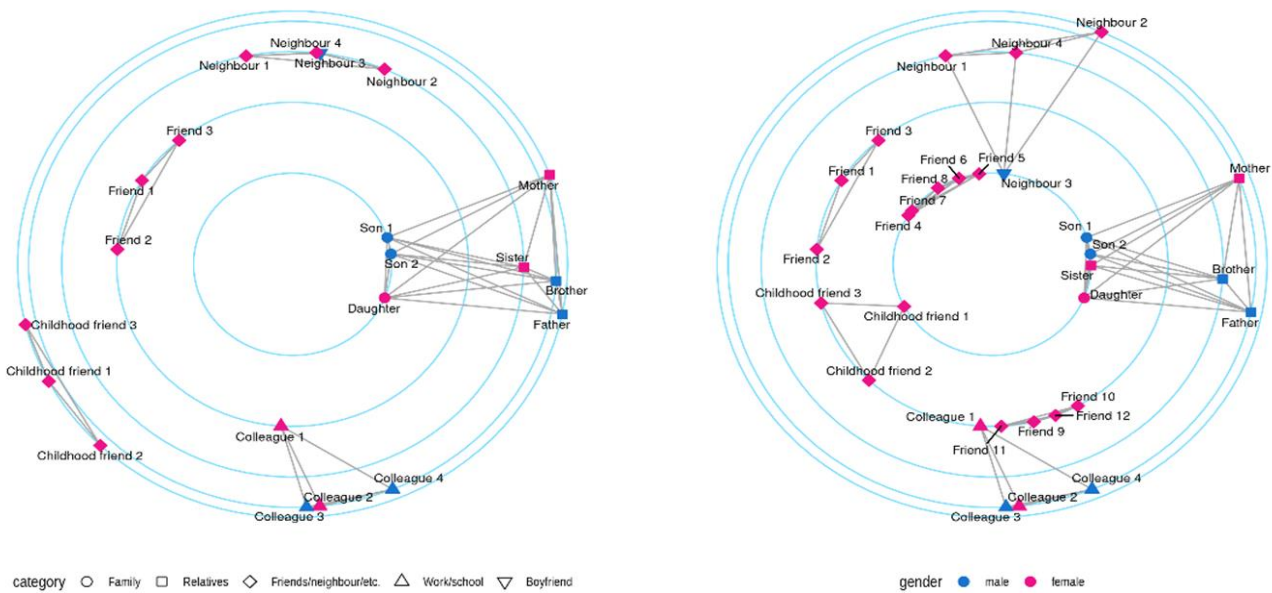

**Figure 1.** Louise's personal networks, representing the people who were around here before and after she called the police.

Using these data, we describe the nature of ambivalent and consistent relationships, discussing how they evolved over time, the role they played in defining the context of domestic violence, their input in shaping the possible pathways out of it, and their embeddedness in the personal networks of abused women. Overall, our analytical strategy can be defined as a qualitative analysis of social network relationships, where the objective is to describe relationships' content and dynamics, and to propose relational typologies in which qualitative materials account for the diversity of individual experiences (Bellotti, 2021).

**4. Ambivalent and Consistent Relationships and Their Roles in Dynamics of Domestic Abuse**

In this section, we report the results of the analysis. Several themes emerged from the narratives describing the dynamics of relationships that victims indicate as salient in the process of leaving the perpetrator and reporting him to the police. We organise the themes by looking at how ambivalent and consistent relationships play a role in supporting women to reframe their relationships as abusive, or in dismissing the violence; in charging perpetrators with responsibilities for the abuse, or in siding with them; in relieving women from role-expectations and obligations, or in burdening them with further responsibilities; and in supporting, or challenging, their pathways out of domestic abuse. In each of these sections, we highlight the alignments and discrepancies between expectations and actions, fleshing out the relational dynamics of ambivalent and consistent ties.

*4.1. Recognising the Violence: When the Network Dismisses or Acknowledges the Abuses*

The first theme highlighted in our analysis relates to the relational dynamics that take place when women first disclose abuse or when the network starts suspecting it and the consequent network responses. In some cases, people around them acknowledged that women were victims of violence and helped them define the situation as domestic abuse; in other cases, they minimised the situation, did not fully believe the victims, or even considered them partially responsible for the abuse.

The example of Nicole is a good illustration of the consistent support that she received from her family in redefining her relationship with her partner as abusive. Nicole had been with the perpetrator for four years, and they had children together. The violence started in the final year of the relationship, and Nicole confided in one of her sisters while she did not initially disclose the abuse to the rest of the family. However, her mother and brother suspected it, and her sister-in-law gave her the number of a service for abused women. Nicole contacted the service, who strongly advised her to call the police: during a violent episode, Nicole remembered the words of the woman from the service and decided to call the police. After she filed a police report, the whole family consistently supported her in leaving the perpetrator's house and finding a permanent home for herself and the children.

In the case of Amelia, however, her sister provided an ambivalent response. Amelia had been with the perpetrator for four years and maintained an amicable relationship after they split up. On one occasion, she invited him to her apartment, but once there, he sexually abused her. After the incident, she immediately called her sister, who, according to Amelia, reacted in an unexpected manner and minimised the abusive incident. Amelia also suspects that when subsequently interviewed by the police, the sister provided an inconsistent account and undermined the investigation:

> My sister did not really want to understand it and did not really want to accept it. She was very shocked, I would say. She could have been supportive, but she ran away from what had happened.

While Amelia's sister's reaction can be defined as negative ambivalent, as she failed to acknowledge the violent situation in a way that aligned with Amelia's expectations, her grandparents' presence in the network can be interpreted as negative consistent. Amelia describes them as people from a different generation who would not understand what had happened and why it was wrong: Amelia could not see the point of disclosing the abuse to them, as she was not expecting any constructive support. In the case of Amelia, her friends were the ones who helped her frame the situation as abusive and ultimately convinced her to call the police.

In some cases, people in the networks are the ones who reframe the situation as violent even when the victim consistently denies the abuse. Martina, for example, was in a relationship with the perpetrator for ten years, with mounting psychological abuse that had turned into physical violence in the final two years, during which time he also progressively isolated her by denying access to social media and spreading malignant gossip about her. Despite the psychological abuse and injuries, Martina acknowledges that while she could see her relationship was destructive, she did not frame it as violent. However, her mother and father suspected that Martina's partner was violent, and they even installed cameras to detect what was going on. While this initially angered Martina, she later recognised, once out of the abusive relationship, that this was a way of supporting her:

> Now, in retrospect, I am very grateful. It had to be done. In the end, it was about life and death; they have saved my life.

In this case, her parents' response was perceived as negative ambivalent at first, as they tried to impose a definition of a violent relationship that Martina did not accept, but it was then reframed as positive consistent once the three of them agreed to reframe the situation as abusive.

These three examples illustrate how on the one hand, social support, when consistent, can help women reframe their relationships as abusive and convince them to report it to the police. However, it also shows how expectations are culturally embedded in generalised roles: a sister, for example, should be supportive, and if she acts differently from what women expect, her support is perceived as ambivalent. Older generations, instead, are not always expected to share the same culturally accepted references and expectations, and so disclosure to them may be withheld or support from them may be ambivalent. Also, the victim and her network need to agree on the modalities in which support is expected and provided (Coohey, 2007).

*4.2. Violent Abuser, but Also Fathers and Partners: When Women and Their Networks See Perpetrators in Different Lights*

A second theme emerging from the qualitative material relates to the way women and their personal networks define the perpetrator: in some cases, they may see him predominantly as an abuser, and therefore accountable for the violence. However, they may also see him as a father with rights toward his children, or as a partner whose violence is attributed to psychological problems or substance addiction, therefore making him less responsible for the abuse.

The case of Olivia is emblematic in this regard. Olivia met the perpetrator when she was in her mid-teens; she was with him for 16 years, during which period they had children together. He was violent from the very beginning of their relationship, but she justified it as him being a heavy drinker and drug abuser. She refrained from disclosing the abuse to her network and the police as she was afraid of losing custody of the children. One day the perpetrator abused her so severely in front of the children that she decided to leave him and report him to the police. The perpetrator was sentenced to a period in prison, and Olivia moved first to a protected accommodation and then to her own flat. After the police report, the perpetrator's siblings turned against her: they insisted on the perpetrator's right to see the children and blamed Olivia for keeping the children away from their father. One of the siblings went as far as filing a complaint to social services accusing Olivia of neglecting the perpetrators' children, a complaint which was dismissed. In this case, the perpetrator's siblings challenged the definition of the perpetrator as a violent person to superimpose it with one of a parent who has rights over his children: we do not directly know if their responses can be seen as negative consistent or negative ambivalent because Olivia does not explain if she was expecting them to react as they did. However, she clearly felt let down by the perpetrators' siblings. Nevertheless, the example shows how perpetrators are not only abusers but also sons, brothers, and fathers, and therefore parts of the network, especially when directly related to him, may take his side.

The situation is different in cases where the victim herself feels somehow responsible for the perpetrator, and therefore is hesitant to leave him. As it was in the case of Elizabeth, whose perception of the abusive relationship swung back and forth from her feeling responsible for the perpetrator and thus wanting to stay and "fix things," to acknowledging the violence and wanting to leave him. Elizabeth had been with the perpetrator for five years, and he started abusing her six months after having moved in together. During the interview, Elizabeth accounts for the perpetrator's violence as he was addicted to drugs, and explains that after each incident, he felt so bad about his behaviour that she eventually felt sorry for him. At the time, Elizabeth was living with the children she had had with her previous husband, who had witnessed the violence and disclosed it to their father: as Elizabeth refused to leave the perpetrator, the children moved in with their father, who informed Elizabeth's family about the abuse. Despite her whole family confronting her, she kept denying the violence, so her whole network eventually withdrew: her family banned the perpetrator from social events, her kids refused to see her, and Elizabeth decided to move out of the suburban area where she lived to hide the perpetrator's violence and his drug problems from friends and neighbours. The violence escalated to the point that Elizabeth got so scared that she moved in with her sister, and the perpetrator started threatening the whole family. When she finally decided to file a police report, the case was dropped by the prosecutor and the perpetrator was not convicted, leaving Elizabeth highly disappointed by the outcome of the legal procedure. In this case, the response of Elizabeth's network was perceived as ambivalent because, to convince her to leave the perpetrator, it completely withdrew its support. Elizabeth, during the interview, acknowledges that it was really important that her family distanced themselves, especially the children because it was ultimately the reason why she decided to leave the perpetrator. She, however, does not talk to them about the perpetrator, as she still has feelings for him, and she finds her family judgmental and patronising:

> They distanced themselves because they loved me, they just wanted him to [disappear], they were terrified that he would kill me.

The process of reframing an intimate relationship in abusive terms, and how the network around the victim may or may not help, empirically shows how disclosure is a fluid progression that needs to develop a consistent narrative for the network to adapt its support and modify responses (Latta & Goodman, 2011). When narratives are not consistent, for example, when part of the network, or the victim herself, superimpose the definition of the perpetrator as a vulnerable person and rightful father over that of a violent partner, it becomes difficult for families or friends to plan a pathway of action to help the victim out of the abusive relationship.

*4.3. Roles and Responsibilities: When Victims Are Also Mothers, Daughters, Sisters, and Friends*

Another theme that emerged from our analysis relates to the direction of support, or its reciprocity. Female victims of abuse are not just the recipients of help, but they also occupy roles in their networks that come with obligations and responsibilities. Because of that, women may refrain from disclosing the abuse in the fear that the perpetrator may retaliate against their networks, because they do not want to leave children behind, or because they do not want to burden friends and families who may also face difficult times in their lives.

Annika's experiences exemplify the case in which, to protect children, the victims progressively falls into a highly isolated situation. Annika describes the perpetrator as very controlling from the very beginning of the relationship, limiting her access to the phone and disapproving of her family and friends. However, it is only when they moved in together that the violence started, he forbade people from visiting her, and she felt so ashamed that she quit her job and deliberately argued with her family so as to avoid contact. Annika said that she did not want to leave the perpetrator in a rush because she felt responsible for his and his ex-wife's daughters who lived with them, and she feared for their safety. She spent a year secretly planning her escape with the consistent help of one of her close friends, her mother, and the perpetrator's ex-wife. The secrecy of the plan made her relationship with her sister ambivalent because while she disclosed the abuse to her, she did not disclose the escape plan. The sister was initially instrumental in putting the victim in contact with a centre for female victims of abuse (positive consistency) but became angry with the victim when she did not leave the perpetrator, while the victim was expecting more empathy from her (negative ambivalence):

> [I wish she] had been a little more understanding, but she is very impulsive and very principled and then she didn't know [about the plan], so I don't really hold it against her.

Unexpected help came from the perpetrator's ex-wife (positive ambivalence), while Annika's mother was consistent in her support the whole time, not giving up on the victim even when she tried to distance herself.

In other instances, we see reluctance on the part of the victims in reporting the abuse for fear that the perpetrator may retaliate against themselves or people in the network. Ronja's brother, for example, took the initiative and called the police: as a result, Ronja was exposed to even more violence, refused to collaborate with the police, and the perpetrator started threatening the brother, who ceased contact with the victim altogether:

> Even if he meant well, I got in trouble for it. I lived under threat and my family's attempts to contact me only resulted in worse physical violence.

Here the ambivalence lies in the timing of the brother's reaction: not being expected by the victim, she did not make use of the brother's initiative, and the lack of concerted effort to involve the authorities resulted in retaliation, more violence, and the brother's withdrawal of support.

Finally, there are cases in which the victims do not disclose the abuse to people in their networks because they know their family or friends have other issues going on in their lives and do not want to burden them with their problems. Amelia refrained from talking to her brother, and Maja to her sister, because they had their own psychological issues to deal with and would not have coped with being confidants. Anne avoided talking to, and eventually ceased contact with, a pair of friends, married to each other, who were already dealing with the husband's suicide attempt: not only were they unable to provide Anne with any help, but they also needed her support—and she was in no condition to provide it. These relationships can all be interpreted as negative consistent because the victims do not obtain any support from them, but they also do not expect it.

In these examples, we see how networks are not simply sources of support, but they come with role expectations and obligations (Connidis & McMullin, 2002; Stryker, 1980). Women may be victims, but they are also mothers, expected to look after their children; they are friends, expected to be there when other friends need them; they are daughters and sisters, responsible for their families. Not only do women need to negotiate role expectations within each of their relationships, but with the network as a whole. The shape of the network may or may not reinforce their definition of the situation (Bott, 1957): family cliques may present similar views, a highly interconnected network may coordinate interventions, and non-overlapping circles may offer alternative views or conflicting obligations (Krackhardt, 1998).

*4.4. Planning and Enacting Pathways of Escape: When Networks Facilitate or Constrain Available Options*

The final theme that we identify in our analysis relates to the way in which personal networks may—or may not—support the decision of the victims to leave the perpetrator, the timing of it, and its modalities. For example, in the cases of Jeanette and Nicole, the perpetrator's family withdrew support when the victim decided to report the partner to the police. The consistent support the perpetrators' families offered during the relationship turned to ambivalent when the formal institutions were involved, as they would have preferred to solve the situation informally and were concerned about the long-term consequences that the perpetrator may have had to face. On the other hand, victims' family members may condition their support on the victim's decision to report the perpetrator to the police, such as Nora's father, who stated that he would not even meet her if she was not prepared to go to the police, and once she did, he refused to go to court with her, contrary to her expectations:

> My father said: "I will not support you unless you report to the police." He was completely focused just on that. And after the police report I asked him if he wanted to come to the trial, which I thought he would because this lawsuit was obviously important to him….That was quite hard, that he did not care anymore, so I have not received any support at all from him.

A common type of support that turns from consistent to ambivalent is that in which people who initially help the victim then become frustrated and withdraw help when the victims repeatedly change their mind and return to the perpetrator. Nike, for example, disclosed the abuse to her uncle and asked for advice. The uncle suggested reporting the abuse to the police and ending the relationship with the perpetrator, but when the niece did not follow the advice, he refused to talk to Nike again. Kerstin's sister, despite being afraid of the perpetrator, offered a safe place for Kerstin to talk and gave her advice, but when Kerstin kept going back to the perpetrator, she became sarcastic and withdrew from the relationship. The ambivalence of these cases is a good example of support that comes with "strings attached" (Bosch & Bergen, 2006; Moe, 2007), a support that, instead of being unequivocally helpful, imposes further obligations on the victims.

When the victim's decision to leave the perpetrator aligns with her network's expectations, we see several cases of positive consistent support. Olivia's boss and her colleagues helped her by changing her shifts and hiding her when the perpetrator showed up at her workplace. Fia's boss initially lent her an apartment, then her parents bought her one and equipped it with surveillance, while one of her friends took the children to live with her while Fia settled in. Similar to the case Annika described, once Ronja decided to leave the perpetrator, she made a detailed plan on how to coordinate her network's support, so her mother, her aunt, her cousin, and a priest all liaised together to help her when she had to go back home (from where she moved in with the perpetrator) and with financial and living arrangements. This type of consistent support is in line with the positive help described in the literature (Edwards et al., 2012; Lempert, 1997; Mahlstedt & Keeny, 1993).

## 5. Conclusions

This article explores the nature of social networks that surround victims of domestic violence. As such, it focuses on the relational dynamics as described by women who have lived in abusive relationships, managed to find their way out, and recall the experience in hindsight. As we did not talk to victims who were currently in abusive relationships, our results do not extend to those situations. One of the traits we discussed about ambivalent and consistent relationships is how they may change over time, especially after the victims call the police and leave the perpetrators, so accounts from women who currently experience domestic violence may be very different, as relational expectations and disclosure may well be affected.

Intentionally, we focus only on informal sources of support, excluding, for example, social services, police, therapists, or support groups that are sometimes included in the personal networks of abused women. We also cannot say anything about the expectations and experiences that networks may have of the victims, therefore limiting our understanding of the development of reciprocal role-expectations. We only touch upon the interesting element of building relational expectations, which may rely on unspoken communication of needs or on culturally generalised role expectations that are implicit in personal relationships. Expectations may also depend on different views on the acceptability of violence against women, beliefs about men's sexual entitlement, and power inequities in relationships, which have been associated with increasing incidences of domestic violence (McCarthy et al., 2018). Further work is needed to focus on individual perceptions of support and what is reciprocally expected in relationships, especially when it comes from institutionalised sources of support.

Following the theoretical framework of ambivalence and consistence we delineated at the beginning of the article, we observe how women describe their networks' responses as disappointing or surprising when support is expected or unexpected. We observe how such responses may change over time and may be endorsed (or not) by others in the networks. In describing these relationships, we analyse how women and their networks come to define the relationship as violent, how they reframe the roles and responsibilities of both victims and perpetrators, take into account the other people that they may feel responsible for, and elaborate pathways of actions that lead them to terminate abusive relationships. Our analysis shows how each of these important aspects are negotiated within their personal networks and how they play a role in offering opportunities and constraints for potential pathways of action. Within the relational dynamics of these networks, alignment and misalignment in defining the abusive situation highlight role expectations as well as reformulations.

In this, we contribute to the theoretical debate on social networks as a source of social support, but also to the discussion of the role of social support in limiting or enhancing social isolation, as we shed an innovative light on why women suffering domestic abuse may end up socially isolated. When perpetrators cut them off from their supportive networks, these networks may respond by coordinating pathways for escape. However, to do so, their definition of the situation as abusive, their perception of women as victims, and their acknowledgement of victims and perpetrators' roles and responsibilities need to be consistent with the definitions and perceptions of the victims as well as everyone else in the network. If not, the victim herself may withdraw from support, or the network may abandon her.

Therefore, social isolation is not the simple result of a lack of support but of the complex dynamics in which support is offered and accepted, dynamics which need to consider which times and modalities work for both victims and their networks. Our results suggest that when people suspect (or victims disclose) domestic violence, to negotiate consistent forms of support, time and effort should be spent to understand what the victim expects from her personal network as well as what her personal network expects from her. When working with victims or in domestic violence prevention, a focus should be on the alarming sign of networks withdrawing support altogether, which can have the valuable consequence of making victims more aware of the dangerous nature of their romantic relationship. Another focus should be on the potentially unexpected sources of support, such as neighbours and colleagues, and victims should be encouraged to disclose beyond their inner circle of family and close friends.

**Acknowledgements**

This study was funded by FORTE, the Swedish Research Council for Health, Working Life and Welfare (grant no. 2016–00987). The Regional Ethical Review (Lund) reviewed and approved the project (no. 2017/1077). We would like to thank David Schoch at the University of Manchester for the visualisation of the personal network in Figure 1.

**Conflict of Interests**

The authors declare no conflict of interest.

**About the Authors**

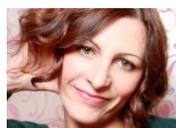

**Elisa Bellotti** is senior lecturer in sociology and member of the Mitchell Centre for Social Network Analysis at the University of Manchester. She has published extensively on applications of social network analysis and mixed methods in substantive sociological fields such as criminal networks, scientific networks, and personal networks. Her recent work focuses on gender aspects of social network formations and outcomes, and health networks. She is the author of *Qualitative Networks. Mixed Methods in Sociological Research* (Routledge, 2015) and co-author of *Social Network Analysis for Egonets* (Sage, 2015).

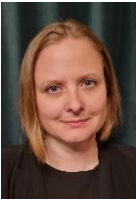 **Susanne Boethius** (PhD) is a researcher at the Department of Sociology at Lund University. Her research interests are interpersonal violence, violence against women, and treatment programs for violent men. Her current research focuses on the social network's involvement and responses to domestic violence.

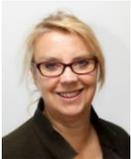 **Malin Åkerström** is professor of sociology at Lund University in Sweden. Her earlier ethnographic studies have concerned social control and deviance. Her most recent book is *Hidden attractions of Administration—The Peculiar Appeal of Meetings and Document* (Routledge, 2021). It is written with co-authors K. Jacobsson, E. Anderrson-Cederholm, and D. Wästerfors, and concerns the involvement and embracement in bureaucratic concerns among human service staff. Apart from this, she has published *Suspicious Gifts—Bribery, Morality, and Professional Ethics* (Transaction/Routledge, 2014), and has published articles in *Sociological Focus*, *Social Problems*, *Symbolic Interaction*, and *Sociological Perspective*, among other journals.

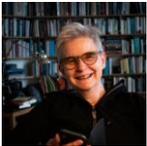 **Margareta Hydén** is professor emerita in social work at Linköping University, Sweden, and visiting professor in criminology at Stockholm University, Sweden. Her research includes interpersonal violence, feminist, and narrative studies. Her recent work focuses on social network's responses to intimate partner violence. She has been acknowledged for moving the research field beyond a mere static victim's perspective by examining the complex ways in which victims negotiate their power, and for developing narrative approaches for the study of sensitive topics.